\documentclass[amsmath,amssymb,pss,preprint,floatfix,superscriptaddress]{revtex4}
\usepackage{graphicx}
\usepackage{stmaryrd}
\usepackage{dcolumn}
\usepackage{bm}
\usepackage{color}
\usepackage{multirow}
\usepackage{times}

\newcommand{\vk}{\ensuremath{\vc{k}}}
\newcommand{\vq}{\ensuremath{\vc{q}}}
\renewcommand{\vr}{\ensuremath{\vc{r}}}

\newcommand{\vR}{\ensuremath{\vc{R}}}
\newcommand{\vX}{\ensuremath{\vc{X}}}
\newcommand{\vx}{\ensuremath{\vc{x}}}
\newcommand{\vchi}{\ensuremath{\vc{\chi}}}
\newcommand{\vv}{\ensuremath{\vc{v}}}

\newcommand{\Pv}{\ensuremath{\boldsymbol{\sigma}}}

\newcommand{\Tmat}{\ensuremath{\mathcal{T}}}

\newcommand{\Ghost}{\ensuremath{G^{\rm host}}}
\newcommand{\Gimp}{\ensuremath{G^{\rm imp}}}
\newcommand{\Gsingle}{\ensuremath{{G_{\rm S}}}}
\newcommand{\Gmulti}{\ensuremath{{G_{\rm M}}}}
\newcommand{\Gs}{\ensuremath{\mathcal{G}}}
\newcommand{\Gsimp}{\ensuremath{\mathcal{G}^{\rm imp}}}
\newcommand{\Gshost}{\ensuremath{\mathcal{G}^{\rm host}}}
\newcommand{\spo}{\ensuremath{\tau}}

\newcommand{\rR}{\ensuremath{\mathcal{R}}}
\newcommand{\rH}{\ensuremath{\mathcal{H}}}
\newcommand{\lR}{\ensuremath{\bar{\mathcal{R}}}}
\newcommand{\lH}{\ensuremath{\bar{\mathcal{H}}}}

\newcommand{\DV}{\ensuremath{\,\Delta V}}

\newcommand{\exJDOS}{\ensuremath{\mathrm{exJDOS}}}

\newcommand{\VBZ}{\ensuremath{\Omega_{\rm BZ}}}
\newcommand{\crvol}{\ensuremath{\Omega_{\rm cryst}}}

\newcommand{\rcvol}{\ensuremath{\Omega_{\rm rec}}}

\newcommand{\volimp}{\ensuremath{\Omega_{\rm imp}}}
\newcommand{\volsub}{\ensuremath{\Omega_{\rm excl}}}
\newcommand{\volscan}{\ensuremath{\Omega_{\rm scan}}}

\renewcommand{\Im}{\mathrm{Im}}

\newcommand{\Tr}{\mathrm{Tr}}

\newcommand{\EF}{\ensuremath{E_{\rm F}}}

\newcommand{\BT}{Bi$_2$Te$_3$}

\newcommand{\klx}{\ensuremath{k_{\parallel}}}

\newcommand{\matr}[1]{\ensuremath{\llbracket #1 \rrbracket}}
\newcommand{\breq}{\nonumber\\&&}

\newcommand{\vc}[1]{\ensuremath{\boldsymbol{#1}}}

\newcommand{\beq}{\begin{equation}}
\newcommand{\eeq}{\end{equation}}
\newcommand{\eq}[1]{Equation~(\ref{#1})}
\newcommand{\lbl}[1]{\label{#1}}

\newcommand{\tit}[1]{\textit{#1}.}
\newcommand{\vol}[1]{\textbf{#1}}
\newcommand{\doi}[1]{{ doi:#1}}

\newcommand{\cita}[1]{\cite{#1}}

\begin{document}

\title{Ab-initio Theory of Fourier-transformed Quasiparticle Interference Maps and Application to the Topological Insulator \BT}

\author{Philipp R\"u{\ss}mann} 
\email{P.Ruessmann@fz-juelich.de}
\affiliation{Peter Gr\"unberg Institut and Institute for Advanced Simulation, 
	Forschungszentrum J\"ulich and JARA, D-52425 J\"ulich, Germany}
\author{Phivos Mavropoulos}\email{Ph.Mavropoulos@fz-juelich.de} 
\affiliation{Department of Physics, National and Kapodistrian University of Athens, GR-15784 Zografou, Athens, Greece}
\author{Stefan Bl\"ugel} 
\affiliation{Peter Gr\"unberg Institut and Institute for Advanced Simulation, 
	Forschungszentrum J\"ulich and JARA, D-52425 J\"ulich, Germany}

\begin{abstract}
	The quasiparticle interference (QPI) technique is a powerful tool that allows to uncover the structure and properties of electronic structure of a material combined with scattering properties of defects at surfaces. Recently this technique has been pivotal in proving the unique properties of the surface state of topological insulators which manifests itself in the absence of backscattering. In this work we derive a Green function based formalism for the \textit{ab initio} computation of Fourier-transformed QPI images. We show the efficiency of our new implementation at the examples of QPI that forms around magnetic and non-magnetic defects at the \BT\ surface. This method allows a deepened understanding of the scattering properties of topologically protected electrons off defects and can be a useful tool in the study of quantum materials in the future.
\end{abstract}

\maketitle

\section{Motivation}

The scanning tunneling microscopy (STM) experiments of Crommie, Lutz and Eigler \cita{Crommie93} 
and Hasegawa and Avouris \cita{Hasegawa93}, revealing
standing density waves of the Cu(111) and Au(111) surface-state
electrons near defects, have pioneered a very powerful, direct method of imaging the surface electron liquid of metals.
Together with Scanning Tunneling Spectroscopy (STS), the method gives
unique insight on the quasiparticle interference (QPI), scattering phase shifts, and lifetime.
Especially when augmented by the Fourier-transformed quasiparticle interference map, 
as proposed by Petersen et al.\ \cita{Petersen98}, the method unveils
scattering properties of quasiparticles off surface defects, giving information
on the scattering vectors among points of the band structure. In this context, Fourier-transformed QPI 
maps provided one of the first experimental proofs of the existence of topological insulators \cita{Roushan09},
because it revealed the \emph{absence} of intensity at back-scattering vectors, just as predicted by theory.

From a theoretical point of view, the calculation of QPI maps has been largely based on model methods, e.g.\ on topological insulator surfaces \cite{Lee09}, where the surface band structure can be approximated by simple model Hamiltonians. In general, however, density-functional based methods are necessitated for a realistic description of the surface electronic structure and in particular of the impurity potential, where the charge relaxation around the impurity plays a major role in the correct description of the scattering phase shifts. A difficulty in density-functional calculations is that the density oscillations induced by the defect are very long-ranged, reaching tens or even hundreds of nanometers, so that supercell methods cannot practically reach this limit. These challenges can only be met by ab-initio Green function embedding methods, like the Korringa-Kohn-Rostoker (KKR) method. 

As an example of an application, we refer to the calculations by Lounis et al.\ \cita{Lounis11} of QPI on Cu(111) and Cu(001) surfaces due to an isolated impurity buried under the surface. These results show that ab-initio calculations of QPI maps  over quite large surface areas are feasible with Green function techniques. However, in the case of Fourier-transformed QPI map, it is practical to express the result directly by a convolution of Green functions \cite{Wang03}, avoiding the intermediate step of calculating the real-space map in a large surface area.

In this paper, we approach this problem and give applications in the field of topological insulators. In Sec.~\ref{sec:Formalism} we outline the formalism for real-space and Fourier-transformed QPI maps within the KKR method. Furthermore, we discuss the Fourier-transformed QPI for the practical case of multiple impurities and argue that the many-impurity problem is well approximated by the single-impurity result. We also discuss the extended joint density of states approach (exJDOS). In Sec.~\ref{sec:Applications} we apply our formalism on the topological insulator \BT\ with surface impurities. This is implemented in the \texttt{JuKKR} code package \cita{jukkr}. Finally, we conclude with a summary in Sec.~\ref{sec:Summary}.

\section{Formalism \lbl{sec:Formalism}}

Within the Tersoff-Hamann approximation \cita{Tersoff85}, the STM differential tunneling conductivity at a bias voltage $U$ is related
to the space- and energy-resolved density of states $n(\vr;E)$ at energy $E$ and at the position \vr\ of the tip: $n(\vr;\EF+eU)\propto \frac{dI}{dU}(U)$ 
(\EF\ is the Fermi level). In a QPI experiment we are interested in the difference of density induced by an impurity with respect to the pristine host surface,
\begin{equation}
\Delta n(\vr;E)= n^{\rm imp}(\vr;E)-n^{\rm host}(\vr;E). 
\lbl{eq:TersoffHamann}
\end{equation}

\subsection{Green function and \Tmat-matrix approach}

The Green function and \Tmat-matrix approach is a well-established method of calculating the density of systems with impurities. It has been applied to the QPI problem in ab-initio and model calculations, e.g.\ in Refs.~\onlinecite{Wang03, Lee09, Guo10}. Here we give the formalism in an explicit real-space representation, because it forms the basis of the formalism and discussion in subsequent sections.

The difference in the local density of states is connected to the one-electron Green function of the system with impurity, $\Gimp(\vr,\vr';E)$, and to the one of the pristine host, $\Ghost(\vr,\vr';E)$,  by the well-known identity 
\begin{equation}
\Delta n(\vr;E)=-\frac{1}{\pi}\Im\Tr\,
\left[ \Gimp(\vr,\vr;E)-\Ghost(\vr,\vr;E)\right].
\lbl{eq:nG}
\end{equation}
The trace is implied with respect to spin indices; in the case of a relativistic formalism with Dirac four-vector states, the trace includes also the large and small components. By virtue of the Dyson equation, $\Gimp-\Ghost=\Ghost\DV\Gimp=\Gimp\DV\Ghost$, the difference in the Green function $\Delta G=\Gimp-\Ghost$ is written as
\begin{eqnarray}
\Delta G(\vr,\vr;E) =\int \!\!d^3r'\!\!\int\!\! d^3r''
\Ghost(\vr,\vr';E)\,\Tmat(\vr',\vr'';E)\,\Ghost(\vr'',\vr;E)
\lbl{eq:DeltaG}
\end{eqnarray}
in terms of the \Tmat-matrix,
\begin{equation}
\Tmat(\vr',\vr'';E) = \DV(\vr')\,\delta(\vr'-\vr'')+\DV(\vr')\,\Gimp(\vr',\vr'';E)\,\DV(\vr'').
\end{equation}
$\DV=V^{\rm imp}-V^{\rm host}$ is the difference in the potential between the impurity and host systems. The \Tmat-matrix has the advantage that it is confined in the small region where the potential difference does not vanish and has to be calculated only once, irrespective of the range of $\vr$ in the QPI calculation. For the calculation of $\Ghost$, the translational invariance of the surface allows us to use the Bloch theorem. Decomposing the position vector as $\vr=\vR+\vx$, where $\vR$ is a lattice translation vector parallel to the surface plane, while $\vx$ is a vector in the primitive cell, we write
\begin{equation}
\Ghost(\vR+\vx,\vR'+\vx';E) = \frac{1}{\rcvol}\int d^2k\, \Ghost_{\vk}(\vx,\vx';E)e^{i\vk\cdot(\vR-\vR')}
\lbl{eq:FTG0}
\end{equation}
where the Fourier-transform of the Green function obeys the spectral representation
\begin{equation}
\Ghost_{\vk}(\vx,\vx';E) =\sum_{\alpha} \frac{\Psi_{\alpha\vk}(\vx)\Psi^{\dagger}_{\alpha\vk}(\vx')}{E-E_{\alpha\vk}+i0}
\lbl{eq:Gsp}
\end{equation}
Here, $\Psi_{\alpha\vk}$ is the host wavefunction, $\alpha$ is the band index, $\rcvol=(2\pi)^2/\crvol$, with $\crvol$ the total crystal surface area, and $i0$ represents an infinitesimal imaginary energy. The difference in Green functions, Equation~(\ref{eq:DeltaG}), takes then the form
\begin{eqnarray}
\Delta G(\vR+\vx,\vR+\vx;E) &=& 
\frac{1}{\rcvol^2} \int d^2k \int d^2k' e^{i(\vk-\vk')\cdot\vR} \,\sum_{\vR',\vR''} \int_{\vR'} \!\!\!d^3x'\!\!\!\int_{\vR''}\!\!\! d^3x''  \, e^{-i\vk\cdot \vR'}e^{i\vk'\cdot\vR''}  \nonumber\\
&\times& 
\Ghost_{\vk}(\vx,\vx';E)\,
\Tmat(\vR'+\vx',\vR''+\vx'';E)\, 
\Ghost_{\vk'}(\vx'',\vx;E)  \lbl{eq:DeltaG1} \\
&=& 
\frac{1}{\rcvol^2} \int \!\!\! d^2k\!\!\! \int\!\!\! d^2k' \sum_{\alpha\alpha'}
\frac{\Psi_{\alpha\vk}(\vx)\Tmat_{\alpha\vk\alpha'\vk'}(E)\Psi^{\dagger}_{\alpha'\vk'}(\vx) }{(E-E_{\alpha\vk}+i0)(E-E_{\alpha\vk'}+i0)}e^{i(\vk-\vk')\cdot\vR}
 \lbl{eq:DeltaG2} 
\end{eqnarray}
The sum over $\vR'$, $\vR''$ and the integration over $d^3x'$, $d^3x''$ is confined to the sites where the \Tmat-matrix (and the impurity perturbation \DV) is non-vanishing. The former expression, Equation~(\ref{eq:DeltaG1}), includes phase factors $e^{-i\vk\cdot \vR'}e^{i\vk'\cdot\vR''}$ for the inter-lattice-site propagation of the host Green function, when the impurity spreads over many lattice sites. The latter expression (\ref{eq:DeltaG2}) is a more compact form where the  matrix elements $\Tmat_{\alpha\vk\alpha'\vk'}(E)=(\Psi_{\alpha\vk},\Tmat(E) \Psi_{\alpha'\vk'})$ were introduced (note that the summation includes all states, not just the ones at energy $E$). It leads to the stationary phase approximation \cite{Lounis11} pinning the energy to the energy-shell $E$, if the observation point is far from the impurity ($|\vR|\rightarrow\infty$).

For the Fourier-transformed QPI we need the Fourier transformation of the Green function along a surface parallel to, and at vertical distance $z$ from, the crystal surface:
\begin{eqnarray}
\Delta G(z;\vq;E) &=& \int_{(z)} d^2r \Delta G(\vr,\vr;E) \, e^{-i\vq\cdot\vr} 
\lbl{eq:FTG1}\\
&=&
\frac{\VBZ}{\rcvol^2}
\!\!\int_{(z)} \!\!d^2x  \, e^{-i\vq\cdot\vx}\int\!\! d^2k\, 
\sum_{\vR'\vR''} e^{-i\vk\cdot\vR'} \,e^{i(\vk-\vq)\cdot\vR''} \nonumber\\
&\times& \!\!
\int_{\vR'} \!\!d^3x' \!\!\int_{\vR''} \!\!d^3x''
\Ghost_{\vk}(\vx,\vx';E)\,
\Tmat(\vR'+\vx',\vR''+\vx'';E)\, 
\Ghost_{\vk-\vq}(\vx'',\vx;E)   
\lbl{eq:FTG2}
\end{eqnarray}
In the step from (\ref{eq:FTG1}) to (\ref{eq:FTG2}) we used Equation~(\ref{eq:DeltaG1}) and we removed  
 lattice sum $\sum_{\vR}$  by virtue of the identity
$\sum_{\vR}e^{i(\vk-\vk'-\vq)\cdot\vR}=\VBZ\delta(\vk-\vk'-\vq)$ (\VBZ\ is the surface Brillouin zone area). The result represents the convolution of two Green functions, as expected from the Fourier transform of their products. Employing Equation~(\ref{eq:nG}), we arrive at the following expression for the Fourier transformed QPI:
\begin{eqnarray}
\Delta n(z;\vq;E) &=& \int_{(z)} d^2r \,n(\vr;E)\, e^{-i\vq\cdot\vr} \nonumber\\
&=& -\frac{1}{2i\pi}
\Tr\,\left[
\Delta G(z;\vq;E) - \Delta G(z;-\vq;E)^*
\right]
\lbl{eq:FTQPI}
\end{eqnarray}
where it is implied that the complex conjugation operation $\Delta G^*$ is done after the Fourier transformation.
This result can easily be generalized for the spin density with $\Pv\Delta G$ in the place of $\Delta G$ (\Pv\ is the vector of Pauli matrices). 

The strongest density change measured by the STM is induced directly ``above the impurity,'' i.e., at a vertical distance $z>0$ from the position where $\DV\neq0$. This region is often excluded in from the Fourier transformation in experiment \cita{SessiPV}, otherwise the image is dominated by the transform of the impurity shape \cita{Beidenkopf2011}, while one seeks the scattering vectors. Additionally, the region close to the impurity is also excluded sometimes, because it may produce spurious background effects in the experiment \cite{Hormandinger94a,Hormandinger94b}. In the calculation, the contribution of the excluded region (indicated by $\volsub$) must be subtracted explicitly, because the form (\ref{eq:FTG2}) already includes a summation over all lattice sites. Thus we define
\begin{eqnarray}
\Delta \breve{n}(z;\vq;E) &=& \Delta n(z;\vq;E) - \int_{\volsub} d^2r\,n(z;\vq;E) \,e^{-i\vq\cdot\vr} \\
&=&
\Delta n(z;\vq;E) + \frac{1}{\pi}\Im\Tr \int_{\volsub} d^2r \,\Delta G(\vr,\vr;E) \,e^{-i\vq\cdot\vr}.
\lbl{eq:denout}
\end{eqnarray}
However, since $\volsub$ is finite-sized, the integration is straightforward in real space.

\subsection{Expression in the KKR formalism}

In the KKR method, the Green function is expanded in site-dependent scattering wavefunctions at sites $n$. The vacuum is also described in a site-centered way by a continuation of the lattice structure beyond the surface, with the corresponding ``empty sites'' containing no atoms but a finite electron density. We denote the general position by $\vr=\vX_n+\vx=\vR_i+\vc{\chi}_\mu+\vx$, where the combined index $n=(i;\mu)$ defines a site $\vX_n$ by the lattice-vector $\vR_i$ and the sub-lattice vector $\vc{\chi}_\mu$, and where $\vx$ is a position vector in the atomic site with respect to the site center.
We employ the regular, $\rR^n_{L}(\vx;E)$, and irregular, $\rH^n_L(\vx;E)$, solutions of the scattering problem of the potential in the vicinity of the site $\vR_{n}$, where $L$ comprises angular momentum and spin indices of the incoming wave. $\rR^n_{L}$ and $\rH^n_L$ are $(2\times1)$ column-vectors in Pauli-Schr\"odinger theory and $(4\times1)$ column vectors in Dirac theory. Also the corresponding left-hand side solutions are needed, denoted by $\lR^n_{L}(\vx;E)$ and $\lH^n_L(\vx;E)$, respectively, that are row-vectors. The expansion breaks the Green function down into a single-site term and a multiple scattering term,
\begin{eqnarray}
G(\vX_n+\vx,\vX_{n'}+\vx';E) 
&=& \sum_L \Gsingle^n_{L}(\vx,\vx';E)\delta_{nn'} + \sum_{LL'}\Gmulti^{nn'}_{LL'}(\vx,\vx';E) \nonumber\\
&=&
-i\kappa\sum_L \big[\rR^{n}_L(\vx;E) \, \lH^{n}_L(\vx';E) \,\theta(x'-x) 
\breq
\ \ \ \ \ \ \ \ \ \ \ \ \ \ \ \ \ \ \ +      \rH^{n}_L(\vx;E)  \lR^n_L(\vx';E)  \,\theta(x-x')   \big]\,\delta_{nn'} \breq
+\sum_{LL'}\rR^{n}_L(\vx;E)\,\Gs^{nn'}_{LL'}(E)\,\lR^{n'}_{L'}(\vx';E)
\lbl{eq:KKRGF}
\end{eqnarray}
where $\kappa=(2mE)^{1/2}/\hbar$ in Pauli-Schr\"odinger theory and $\kappa=(2mE+E^2/c^2)^{1/2}/\hbar$ in Dirac theory. The single-site term describes the Green function of the potential at site $n$  embedded in free space. It only depends on the local potential and its contribution to $\Delta G$ vanishes outside the impurity region. The second term describes the multiple scattering over all sites, expressed by the structural Green function coefficients $\Gs^{nn'}_{LL'}(E)$. These form a matrix $\matr{\Gs(E)}$ that obeys an algebraic Dyson equation. This reads for the host system
\begin{equation}
\matr{\Gshost(\vk;E)}=\matr{g(\vk;E)} + \matr{g(\vk;E)} \matr{t^{\rm host}(E)}\matr{\Gshost(\vk;E)},
\lbl{eq:KKRDyson}
\end{equation}
where we have expressed everything in reciprocal space. $\matr{g(\vk;E)}$ contains the structural Green function coefficients of free space and $\matr{t^{\rm host}(E)}$ is a site-diagonal (\vk-independent) matrix containing the \Tmat-matrices of each site with respect to free space, $t^{{\rm host};\mu}_{LL'}(E)$, expressed in an angular momentum and spin basis. The lattice-part of the Fourier transformation affects only the structural Green functions $\matr{\Gshost(\vk;E)}$, not the \Tmat-matrices or local scattering solutions.

The analogon of the \Tmat-matrix in the KKR method is the \emph{scattering path operator} of the impurity with respect to the host, $\matr{\spo(E)}$. It is expressed in terms of the single-site \Tmat-matrices of impurity and host, $\Delta t^n_{LL'}(E)=t^{{\rm imp};n}_{LL'}(E)-t^{{\rm host};n}_{LL'}(E)$, and of the host structural Green function, by the Dyson-type equation $\matr{\spo}=\matr{\Delta t} + \matr{\Delta t} \matr{\Gshost}\matr{\spo}$. It is not site-diagonal, and has non-vanishing elements $\spo^{nn'}_{LL'}(E)$ only between sites $(n,n')$ for which $\Delta t^n\neq0$ and $\Delta t^{n'}\neq0$. The structural Green function of the system with impurity is then expressed by
\begin{eqnarray}
{\Gsimp}^{nn'}_{LL'}(E) &=& {\Gshost}^{nn'}_{LL'}(E) +
\sum_{n''n'''}\sum_{L''L'''} {\Gshost}^{nn''}_{LL''}(E) \,\spo^{n''n'''}_{L''L'''}(E)\,
{\Gshost}^{n'''n'}_{L'''L'}(E)
\lbl{eq:spo2}
\end{eqnarray}
in analogy to Equation~(\ref{eq:DeltaG}).

Since the vacuum region is geometrically described by layers of empty sites parallel to the crystal surface,
it is convenient to approximate the Fourier integration over a surface at distance $z$ by an integration over a vacuum layer of volume $\volscan$, centered at $z$: $\int d^2r \rightarrow \sum_{n\in\volscan}\int_n d^3r$. 
Expression (\ref{eq:FTG2}) then becomes
\begin{eqnarray}
\Delta G(\vq;E)&=& \Delta \Gsingle(\vq;E) + \Delta \Gmulti(\vq;E), \ \ \ \ \ \ \ \ \ \ \ \ \ \ 
\text{with} 
\lbl{eq:KKRQPIFT}\\
\Delta \Gsingle(\vq;E)&=&
\sum^{\volscan\cap\,\volimp}_{j\nu}e^{-i\vq\cdot(\vR_j+\vchi_{\nu})}
\int_{j\nu} d^3x e^{-i\vq\cdot\vx} 
\,\sum_L \Delta G_{{\rm S};L}^{j\nu}(\vx,\vx;E)
\nonumber\\
 \Delta \Gmulti(\vq;E)&=&
\frac{\VBZ}{\rcvol^2}\sum_{\nu} e^{i\vq\cdot\vchi_\nu}
\int d^2k \sum_{i\mu,i'\mu'}^{\volimp}e^{-i\vk\cdot \vR_{i}} e^{-i(\vk-\vq)\cdot \vR_{i'}} 
\breq\times
 \sum_{LL'L''L'''}{\Gshost}^{\nu\mu'}_{LL''}(\vk;E) \spo^{i\mu;i'\mu'}_{L''L'''}(E)
{\Gshost}^{\mu'\nu}_{L'''L'}(\vk-\vq;E)
\breq\times
\int_{\nu} d^3x\, e^{-i\vq\cdot\vx} 
\, \left[\rR^{\nu}_{L}(\vx;E) \lR^{\nu}_{L'}(\vx;E)\right]
\nonumber
\end{eqnarray}
where we set $n=(j,\nu)$, $\Delta G_{{\rm S};L}^{j\nu}(\vx,\vx;E) = G_{{\rm S};L}^{{\rm imp};j\nu}(\vx,\vx;E) - G_{{\rm S};L}^{{\rm host};j\nu}(\vx,\vx;E)$ is the difference of the single-site part of the Green function between the impurity and the host system, and is taken only in the impurity region \volimp\ (it vanishes outside). In the above expression, the terms $\left[\rR^{\nu}_{L}(\vx;E) \lR^{\nu}_{L'}(\vx;E)\right]$ and $\Delta G_{{\rm S};L}^{i\nu}(\vx,\vx;E)$ are $2\times2$ or $4\times4$ matrices (depending if the Pauli-Schr\"odinger or the Dirac theory is used) and must be traced to form the density [see Equation~\eqref{eq:nG}]. Conveniently, they show no $\vk$-dependence and thus must be calculated only once at each energy; the same is true for the matrix elements of the scattering path operator, $\spo^{i\mu;i'\mu'}_{L'L''}(E)$. The only quantities that need to be calculated for a dense set of $\vk$-points (which implies a large numerical effort) are the host structural Green functions (Equation~\ref{eq:KKRDyson}). Fortunately, by virtue of the principal layer and decimation techniques \cite{Godfrin91,Wildberger97,Sancho85}, the latter can be computed with a numerical effort that grows linearly with the number of atomic layers in the film, making possible the accurate simulation of the QPI in thick films (of the order of hundreds of atomic layers, if necessary) or semi-infinite geometries. 

If we wish to calculate the quantity $\Delta \breve{n}(z;\vq;E)$ (Equation~\ref{eq:denout}), i.e., exclude the impurity and its immediate surroundings (indicated by $\volsub$ in Equation~\ref{eq:denout}) from the Fourier transformation, then Equation~(\ref{eq:KKRQPIFT}) changes. The single site term $\Delta \Gsingle$, vanishes automatically outside \volimp. However, we must also explicitly subtract the contribution of the multiple-scattering term in $\volsub$.  The result is given by replacing $\Delta \Gsingle(\vq;E)$ by the following correction to the multiple scattering part
\begin{eqnarray}
\mathcal{C}_{\rm M}(\vq;E) &=& 
\sum^{\volsub}_{j\nu}
e^{-i\vq\cdot(\vR_j+\vchi_{\nu})} 
\sum^{\volimp}_{i\mu,i'\mu'}\sum_{LL'L''L'''}{\Gshost}^{j\nu;i\mu}_{LL''}(E)
\,\spo^{i\mu,i'\mu'}_{L''L'''}
\,{\Gshost}^{i'\mu';j\nu}_{L'''L'}(E)
\breq\times
\int_{i\nu} d^3x \,e^{-i\vq\cdot\vx}
\left[\rR^{\nu}_{L}(\vx;E) \lR^{\nu}_{L'}(\vx;E)\right]
\end{eqnarray}

For the calculation of the integrals $\int_{\nu} d^3x \,e^{-i\vq\cdot\vx}
\left[\rR^{\nu}_{L}(\vx;E) \lR^{\nu}_{L'}(\vx;E)\right]$
we expand the wavefunctions in spherical harmonics, as is normally done in the KKR method \cite{Papanikolaou02}, and we do the same for the exponential by the identity $e^{-i\vq\cdot\vx}=4\pi\sum_{lm}i^l j_l(qx) Y_{lm}(\vk)Y^*_{lm}(\vx)$, where $Y_{lm}$ are spherical harmonics. Thus the integral is decomposed in a spherical and an angular part. The integrals containing irregular functions that contribute to $\Delta \Gsingle_L(\vq;E)$ are handled in an analogous way.

In summary, in the KKR method we calculate the quantities $\Delta n(\vq;E)$ (Equation~\ref{eq:FTQPI}) and $\Delta \breve{n}(\vq;E)$ (Equation~\ref{eq:denout}) by:
\begin{eqnarray}
\Delta n(\vq;E) &=& -\frac{1}{2i\pi}\Tr\left[\Delta \Gmulti(\vq;E)-\Delta \Gmulti(-\vq;E)^* +\Delta\Gsingle(\vq;E) -\Delta\Gsingle(-\vq;E)^*\right]\\
\Delta \breve{n}(\vq;E) &=& \Delta n(\vq;E)+\frac{1}{2i\pi}\Tr\left[\mathcal{C}_{\rm M}(\vq;E)-\mathcal{C}_{\rm M}(-\vq;E)^*\right]
\end{eqnarray}

\subsection{Multiple scattering among impurities}

The experimental QPI Fourier transform is usually performed over a large surface area comprising many impurities. A direct simulation of this experiment should account for the contribution of the multiple-scattering events between impurities to the QPI. However, this is numerically expensive, since it involves the calculation of a large \Tmat-matrix, corresponding to the collection of all impurities in a large supercell, and perhaps even a statistical average over many impurity configurations. Fortunately,  the Fourier-transformed QPI of a single impurity is an excellent approximation to the result of a random impurity distribution, simplifying the calculations. Fang et al.~[\onlinecite{Fang13}] have shown this approximation to hold to lowest order in the potential difference \DV, i.e., in the Born approximation to the scattering amplitude. Here we argue that the approximation holds in general, allowing for the treatment of strong, e.g., resonant, scattering.

First we discuss the form of the multiple-scattering \Tmat-matrix.
Let $t_n(E)$ be the \Tmat-matrix of a single impurity at site $n$ and expressed in a matrix form in a localized basis set. Then, the full \Tmat-matrix of a collection of impurities obeys the expansion \cite{Rodberg}
\begin{eqnarray}
\Tmat_{nn'}&=&t_{n} \delta_{nn'} 
+ t_n\,\Ghost_{nn'}\, (1-\delta_{nn'}) \,t_{n'}+
\sum_{m\neq n,n'}t_{n}\, \Ghost_{nm}\,t_{m}\,\Ghost_{mn'}\,t_{n'}  +\cdots \\
&=&
t_n\delta_{nn'} +t_n \sum_m \breve{G}_{nm} \,\Tmat_{mn'}. \lbl{eq:mstmat}
\end{eqnarray}
where the matrix $\breve{G}_{nm}=\Ghost_{nm}(1-\delta_{nm})$ contains the site-off-diagonal  part of the Green function (a proof is given in the Appendix). 
The above expression includes all multiple-scattering events among impurities, while avoiding sequential scattering off the same impurity (since $t_n$ contains the sequential site-diagonal scattering to all orders of \DV). 

We assume that the impurities are non-overlapping (which is a
reasonable approximation at low concentration) and identical and are
thus characterized by the same matrix $t_n=t$ $\forall n$. We also
apply in part the stationary phase approximation for the host Green
function \cita{Lounis11}, which is valid at long distances. This is
justified at low impurity concentrations since the largest part of the
surface, where the Fourier transform is performed, is covered by  host
atoms and is far from the impurities. Within this approximation, the
host Green function may be approximated by
$\Ghost(\vR_n+\vx,\vR_{n'}+\vx';E)\approx
K_{nn'}(\vx,\vx';E)\,e^{i\vk_{ nn'}\cdot\vR_{nn'}}$, where $\vk_{nn'}$
is a stationary point on the constant energy surface
$E_{\vk_{nn'}}=E$, and is defined by the property that the group
velocity $\vv_{\vk_{nn'}}$ must be parallel to the vector
$\vR_{nn'}=\vR_n-\vR_{n'}$. The quantity $K_{nn'}(\vx,\vx';E)$
contains the rest of the Green function, including a power-law decay
with distance ($K_{nn'}\propto|\vR_{nn'}|^{-1/2}$ in two
dimensions). The important consequence of this approximation for our
purposes is that the phase of the long-distance propagation,
$e^{i\vk_{ nn'}\cdot\vR_{nn'}}$, is governed only by the stationary
point (in the case of multiple stationary points, a summation over the
corresponding contributions is implied). Then we argue that, in the
Fourier transform of Equation~(\ref{eq:FTG2}),
 the contribution of the first term of the
rhs of Equation~(\ref{eq:mstmat}) is dominant and equal to the single-impurity contribution, while
the remainder (the contribution of $t\,\breve{G}\, \Tmat$) is negligible.

We decompose the Green function difference [Equation~(\ref{eq:DeltaG})] in two terms corresponding to the decomposition of the \Tmat-matrix (\ref{eq:mstmat}):  
\begin{eqnarray}
\Delta G_{nn}&=&\Delta G^{(1)}_{nn}+\Delta G^{(2)}_{nn} \\
&=&\sum_m\Ghost_{nm}\,t\,\Ghost_{mn} +\sum_m \Ghost_{nm}\,t\sum_{n'n''} \breve{G}_{mn'}\Tmat_{n'n''}\Ghost_{n''n}.
\lbl{eq:ms2}
\end{eqnarray}
All first-order terms, $\Ghost_{nm}\,t\,\Ghost_{mn}$, give identical contributions to the Fourier transform $\Delta G(\vq)$, because both $\Ghost_{mn}$ and $\Ghost_{nm}$ depend on the sites $n$ and $m$ only via the difference $\vR_{nm}$. If $N$ is the number of impurities, and setting one impurity at position $m=0$, we have $\sum_n\int_n d^2 x e^{-i\vq\cdot(\vx+\vR_n)} \sum_{m} \Ghost_{nm}\,t\,\Ghost_{mn}=N\sum_n\int_n d^2 x e^{-i\vq\cdot(\vx+\vR_n)} \Ghost_{n0}\,t\,\Ghost_{0n}$, which can be shown by changing the  summation over $\vR_m$ to $\vR_{nm}$. This is, however, not true for the higher-order terms. Applying the stationary phase approximation to $\Ghost_{n''n}$ in Equation~(\ref{eq:ms2}) (last term), we obtain $\Delta G^{(2)}_{nn}\approx \sum_m \Ghost_{nm}\,t\sum_{n'n''} \breve{G}_{mn'}\Tmat_{n'n''}K_{n''n}e^{i\vk_{n''n}\cdot\vR_{n''n}}$. In the Fourier transformation, the translational symmetry of the host allows us again to place $m=0$ by a re-indexing, but the contribution of the last phase, $e^{i\vk_{n''n}\cdot\vR_{n''n}}$, cannot be lifted. Since the impurities are randomly placed, the total contribution of the random phases over all impurities practically cancels in the Fourier transform. Of course, an exact cancellation requires a sum over all configurations and thus cannot take place unless the scanned surface area is infinitely large, which is never the case. However, our analysis shows that the single-site term should always give the dominant contribution. In this respect, a calculation of the single-impurity Fourier transform should give a qualitatively and quantitatively representative picture of the full problem, which is numerically advantageous in an ab-initio calculation.

We tested our hypothesis in a model system of a free-electron surface with $s$-wave-scattering point defects, randomly placed and averaged over 50 configurations. Numerical simulations (not shown here) of 1000 impurities in a $1000\times1000$\AA$^2$ box show that, in the Fourier transform, the single-site term dominates over the multiple-scattering contribution by an order of magnitude.

\subsection{Joint density of states: Ad-hoc model or approximation?}

A frequently used approach to the Fourier-transformed QPI is the Joint Density of States (JDOS) \cite{Hoffmann02, Wang03, Simon11, Roushan09} or extended JDOS (exJDOS) \cite{Sessi16} approach, which is applied if the constant-energy contours $\{\vk|E_{\vk}=E\}$ at energy $E$ are known (e.g. from calculations or from angular-resolved photoemission experiments), but the full Green function $\Ghost$ or the full \Tmat-matrix are not known. Motivated by Equation~(\ref{eq:FTG2}), one defines the quantity 
\begin{eqnarray}
\exJDOS(\vq;E)
&=&\int_{E_{\vk}=E} dk \,n_{\rm surf}(\vk;E)\,M_{\vk,\vk-\vq}\gamma^{\rm STM}_{\vk,\vk-\vq}\,n_{\rm surf}(\vk-\vq;E)
\lbl{eq:exJDOS}
\end{eqnarray}
which a weighted convolution of the spectral amplitude at the pristine crystalline surface, $n_{\rm surf}(\vk;E)=\int_{\rm surf} |\Psi_{\vk}(\vr)|^2\,d^3 r$ (the integration over $r$ takes place in the surface and/or in the vacuum region where the STM is positioned). Here, $\vk$ and $\vk-\vq$ are confined to the constant-energy contour by assumption. The matrix element $M_{\vk,\vk-\vq}$ contains information on the scattering properties of the defect. For topological insulators, where spin-flip scattering is at the center of interest, and for non-magnetic defects, where spin-flip scattering is suppressed, the reasonable approximation $M_{\vk,\vk-\vq}=|\Tmat_{\vk,\vk'}|^2\propto 1+\cos(\vc{s}_{\vk},\vc{s}_{\vk'})$ has been proposed [\onlinecite{Roushan09}], where $\vc{s}_{\vk}$ is the spin polarization vector of the state at $\vk$.  Additionally, a factor $\gamma^{\rm STM}_{\vk,\vk-\vq}=1-\cos(\vv_{\vk},\vv_{\vk-\vq})$ was introduced in Ref.~[\onlinecite{Sessi16}] in order to promote standing wave formation by back-scattering (opposite group velocities), in the spirit of the stationary phase approximation. At the end, $\exJDOS(\vq;E)$ is expected to approximately reproduce $|\Delta n(\vq;E)|$, since both should peak at the scattering vectors.

The JDOS approach has been introduced as an ad-hoc model. The question is if it also constitutes an approximation to the theory expressed by Eqs.~(\ref{eq:DeltaG1}), (\ref{eq:FTG2}), and (\ref{eq:FTQPI}). 
We find that $\exJDOS(\vq;E)\propto|\Delta n(\vq;E)|^2$, under a
number of assumptions that are scrutinized in the following. From Eqs.~(\ref{eq:FTG2}) and (\ref{eq:FTQPI}) we have (dropping the variables $E$ and $z$)
\begin{eqnarray}
|\Delta n(\vq)|^2 &=& \Delta n(\vq) \, \Delta n(-\vq) \nonumber\\
&=& -\frac{1}{4\pi^2}\Tr [\Delta G(\vq) - \Delta G(-\vq)^*] \, \Tr [\Delta G(-\vq)-\Delta G(\vq)^*]
\lbl{eq:Dnq2}
\end{eqnarray}

We employ the stationary phase approximation to the Green function, according to which
\begin{eqnarray}
\Delta G(\vr,\vr;E) &\approx&
-\frac{4\pi^3}{\rcvol^2\,\hbar^2}
\sum_{\vk(\vr),\bar{\vk}(\vr)}
u_{\vk}(\vr)\, \Tmat_{\vk,\bar{\vk}}\,
u_{\bar{\vk}}^\dagger(\vr)                              \nonumber\\
&&\times
\exp\left\{-\frac{i\pi}{4}\left[\text{sign}\left(\frac{\partial^2
	E_{\vk}}{\partial \klx^2}\right)+\text{sign}\left(\frac{\partial^2
	E_{\bar{\vk}}}{\partial {\bar{k}_{\parallel}}^2}\right)\right]\right\}   \nonumber\\
&&\times
(|\vv_{\vk}||\vv_{\bar{\vk}}|)^{-1/2} 
\left|\frac{\partial^2 E_{\vk}}{\partial \klx^2}
\frac{\partial^2 E_{\bar{\vk}}}{\partial {\bar{k}_\parallel}^2}\right|^{-1/2}
\,
\frac{e^{i(\vk-\bar{\vk})\cdot\vr}}{|\vr|}
\lbl{eq:spag}\\
&=& \sum_{\vk\bar{\vk}}a_{\vk}a_{\bar{\vk}} u_{\vk}(\vr)\, \Tmat_{\vk,\bar{\vk}}\,
u_{\bar{\vk}}^\dagger(\vr)    \,e^{i(\vk-\bar{\vk})\cdot\vr}/|\vr|
\lbl{eq:FT2a}
\end{eqnarray}
at large distances $|\vr|$ from the impurity (which is placed at
$\vr=0$). We omitted the band index $\alpha$ in order to simplify the
notation. The discrete summation for a given $\vr$ in (\ref{eq:spag}) runs over
$\vk$-points that are stationary with respect to the host Green function phase,
i.e., they are pinned with respect to the energy,
$E_{\vk}=E_{\bar{\vk}}=E$, and also pinned at such positions on the
constant energy contour, that the group velocity $\vv_{\vk}$ is in the
direction $\vr$ and the group velocity $\vv_{\bar{\vk}}$ is in the
opposite direction, $-\vr$. The symbols $u_{\vk}$ and $u_{\bar{\vk}}$ stand for the lattice-periodic part of the wave-function, while $k_{\parallel}$ and $\bar{k}_{\parallel}$ run on the constant-energy contour. The last expression (\ref{eq:FT2a}) is a convenient abbreviation with obvious shorthand notation for $a_{\vk}$ and $a_{\bar{\vk}}$. The Fourier transformation reads
\begin{eqnarray}
\Delta G(\vq;E) &=& \int d^2r \, \sum_{\vk\bar{\vk}}a_{\vk}a_{\bar{\vk}} u_{\vk}(\vr)\, \Tmat_{\vk,\bar{\vk}}\,
u_{\bar{\vk}}^\dagger(\vr)    \,e^{i(\vk-\bar{\vk}-\vq)\cdot\vr}/|\vr| \\
&=& 
\int_0^\infty dr \sum_{\rm CEC}\int_{\rm CEC} dk_{\parallel} \,\mathcal{D}(k_{\parallel}) \sum_{\bar{\vk}}a_{\vk}a_{\bar{\vk}} u_{\vk}(\vr)\, \Tmat_{\vk,\bar{\vk}}\,
u_{\bar{\vk}}^\dagger(\vr)    \,e^{i(\vk-\bar{\vk}-\vq)\cdot\vr}. 
\lbl{eq:FT3}
\end{eqnarray}
In the last step we converted the integration variable from $d^2r$ to
$rdrd\theta$ and subsequently $d\theta$ to
$\mathcal{D}(k_{\parallel})dk_{\parallel}$, where the variable
$k_{\parallel}$ is a parameter running over all constant-energy contours (CEC) as
$\theta$ forms the unit circle (to each stationary point $\vk$
corresponds a direction $\theta$). The quantity
$\mathcal{D}(k_{\parallel})$ is the integration weight corresponding
to the latter transformation and depends on the exact shape of the
constant-energy contour. The stationary points $\vk$ and $\bar{\vk}$
are now functions of $k_{\parallel}$ on the constant-energy contour,
instead of $\theta$  (essentially $\vk$ coincides with $k_{\parallel}$
on the constant-energy contour). To each $\vk$  there may correspond
multiple $\bar{\vk}$ points of opposite group velocity, therefore the summation over multiple possible
$\bar{\vk}$ for each $k_{\parallel}$ remains.

So far we have only used the stationary phase approximation. In order
to derive the JDOS or the exJDOS model, we must make additional
assumptions. First, the weight $\mathcal{D}(k_{\parallel})$ must be dropped, or set to a constant, since it does not appear in the exJDOS expression. But this is justifiable only when the constant-energy contour is approximately isotropic. Then, the assumption must be made that the dominant contribution to the Fourier transform of Equation~(\ref{eq:FT3}) comes from the points where the phase vanishes ($\vk-\bar{\vk}-\vq=0$), dropping the $r$-integration and confining the $k_{\parallel}$-integration only to the points satisfying the latter condition. In addition, on forming products of the type $G(\vq;E)G(\vq;E)^*$ (and similar) that occur in Equation~(\ref{eq:Dnq2}), products of wavefunctions of the type $u^{\dagger}_{\vk}u_{\vk'}$ expressing the density, as well as products of \Tmat-matrix elements expressing the transition rate, will appear. In order to comply with the JDOS (\ref{eq:exJDOS}), the mixed-$\vk$ (i.e., $\vk\neq\vk'$) density terms must be dropped. The terms $a_{\vk}$ and $a_{\bar{\vk}}$ should be also ignored (or set to a constant). Finally, the weighting factor $\gamma^{\rm STM}_{\vk,\vk-\vq}=\delta(\frac{\vv_{\vk}}{|\vv_{\vk}|}+\frac{\vv_{\vk-\vq}}{|\vv_{\vk-\vq}|})$ [instead of the milder $\gamma^{\rm STM}_{\vk,\vk-\vq}=1-\cos(\vv_{\vk},\vv_{\vk-\vq})$] should be set in the definition of the exJDOS (\ref{eq:exJDOS}) in order to account for the stationary phase approximation.

The above discussion shows that the JDOS and exJDOS approaches are not
quantitative approximations but qualitative models. Still, they
comprise essential parts of the information that one usually seeks in Fourier-transformed QPI spectra and therefore constitute a useful tool for their analysis. 

\section{Applications \lbl{sec:Applications}}

To showcase the use of our newly developed method we apply it to the topological insulator \BT, which hosts nontrivial surface states characterized by spin-momentum locking that are protected by time-reversal symmetry against backscattering.

\begin{figure*}
	\centering
	\includegraphics[width=0.90\linewidth]{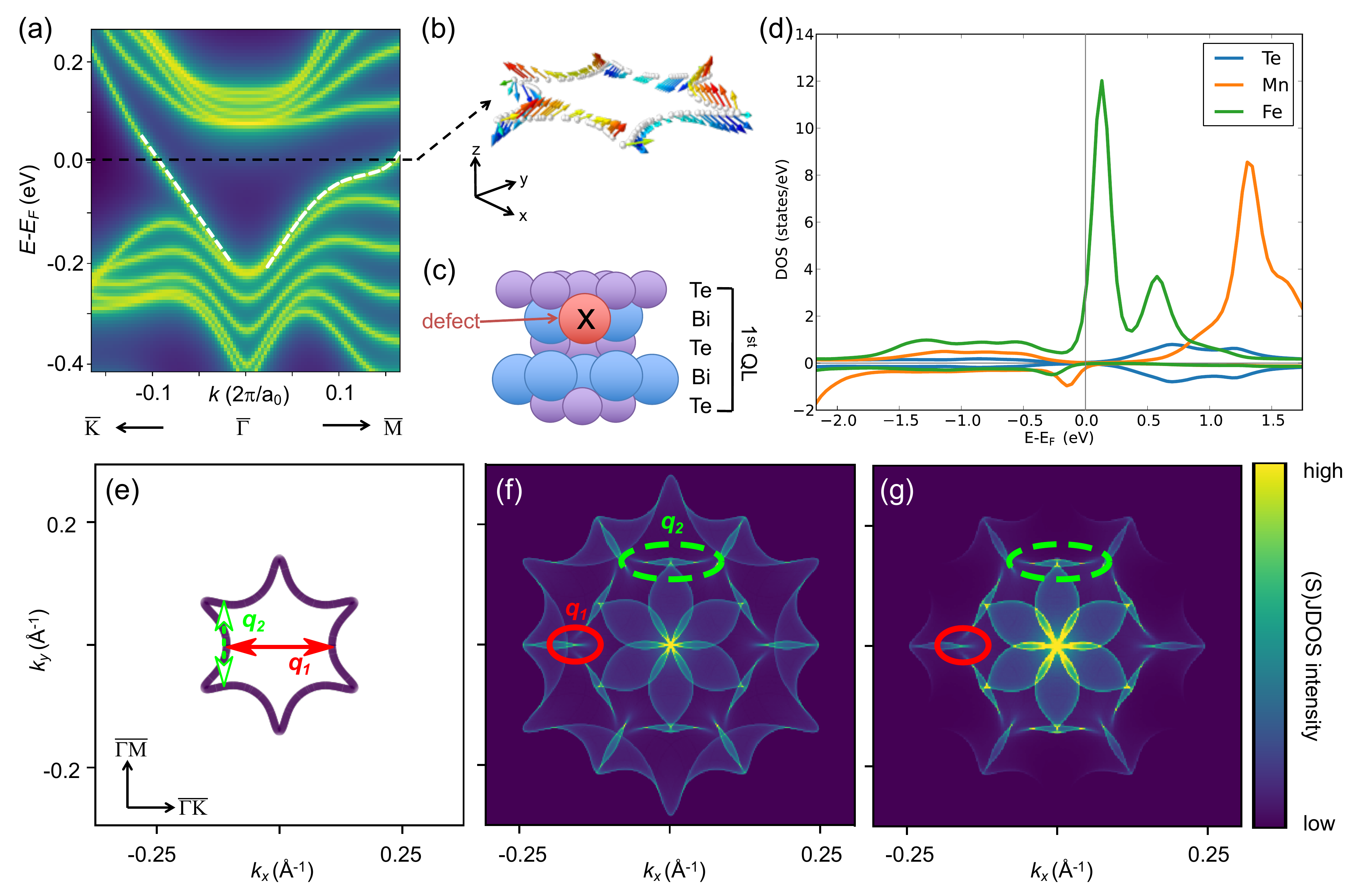}
	\caption{Setting of the numerical simulations. (a)
          Bandstructure of Bi$_2$Te$_3$ with the topological surface
          state highlighted with white dashed lines and (b) its spin
          polarization on the Fermi surface. (c) Location of the
          substitutional impurities in the outer most Bi layer and (d)
          the density of states (DOS) of the different impurities under
          consideration (the positive DOS axis corresponds to spin-up,
          the negative to
          spin-down). (e) Hexagonally warped Fermi surface of Bi$_2$Te$_3$ with the most prominent scattering channels, $\vc{q_1}$ and $\vc{q_2}$, indicated by solid red and dashed green arrows, respectively. (f) JDOS and (g) SJDOS show the conventional way of interpreting FT-QPI images. 
	Image adapted from Ref.~\onlinecite{RuessmannPhD}.}
	\label{fig:figure1}
\end{figure*}

In our density functional calculations within the relativistic full-potential KKR Green function framework \cite{Heers2011, Bauer2013, Long2014, Zimmermann2016} we considered a 6 quintuple layer thick film of \BT\ using the experimental lattice constant \cite{Nakajima1963}, which was chosen such that the ``top'' and ``bottom'' surface states of the thin film decouple. We used an angular momentum cutoff of $\ell_{\mathrm{max}}=3$ including corrections for the exact shape of the cells \cite{Stefanou1990,Stefanou1991} and the local spin density approximation \cite{Vosko1980} for the exchange-correlation functional. The Fermi level was set such that it resides inside the bulk band gap, which ensures that the Fermi surface consists of the topological surface state alone without projections of bulk bands. Such a situation can be achieved experimentally by doping or gating. Figure~\ref{fig:figure1} summarizes the setup of the calculation, where the bandstructure (a) and Fermi surface of the system at hand (b) are shown together with the spin-polarization of the topological surface state.

Single substitutional impurities were embedded into the \BT\ host at
the outermost Bi site as indicated in Figure~\ref{fig:figure1}(c). We
considered non-magnetic $\mathrm{Te}_\mathrm{Bi}$ (the subscript
indicates the substituted position) defect, which occurs naturally as
an anti-site defect, as well as magnetic $\mathrm{Mn}_\mathrm{Bi}$ and
$\mathrm{Fe}_\mathrm{Bi}$ impurity atoms. It has been previously shown \cita{Hor2010,Watson2013,Zhang2013} 
that the substitutional Bi site is a thermodynamically stable position of transition metals in \BT, indicating the relevance of our findings. 
The impurities were embedded into the host system self-consistently making use of the Dyson equation \cite{Bauer2013} while neglecting structural relaxations around the impurities.  The first shell of nearest neighbours was included in the calculation for a correct screening of the charge of the impurities. The resulting density of states for the three defects are shown in Figure~\ref{fig:figure1}(d). 
While the density of states at the Fermi level is small for the $\mathrm{Te}_\mathrm{Bi}$ defect, the incompletely filled $d$-shell of the transition metal impurities results in a higher density of states at the Fermi level. In accordance to Hund's rule, we find that the $\mathrm{Mn}_\mathrm{Bi}$ defect is close to  half-filling whereas the Fermi level bisects the $d$-state resonance of the $\mathrm{Fe}_\mathrm{Bi}$ impurity.
From analyzing the impurity density of states and in the spirit of the Friedel sum rule, which was recently demonstrated to hold in this class of materials \cite{Ruessmann2017}, one expects considerable differences in the scattering properties of the different impurities. In addition, the magnetic nature of the Mn and Fe defects is expected to reopen the forbidden backscattering channel due to breaking of time-reversal symmetry.

The strong hexagonal warping in \BT\ leads to a snowflake-like shape of the Fermi surface that enables two major scattering channels \cita{Fu2009, Beidenkopf2011}, $\vc{q_1}$ and $\vc{q_2}$, which are depicted in Figure~\ref{fig:figure1}(e). The backscattering channel ($\vc{q_1}$) is expected to be suppressed by time-reversal symmetry for scattering off non-magnetic defects and can be re-opened by scattering off magnetic defects. The trivial scattering channel $\vc{q_2}$ is however always possible. These facts are illustrated by the JDOS [Eq.\eqref{eq:exJDOS} with $M_{\vk,\vk'}=\gamma^{\rm STM}_{\vk,\vk'}=1$] and spin-conserving JDOS (SJDOS, $M_{\vk,\vk'}=\vc{s}_{\vk}\cdot\vc{s}_{\vk'}$, $\gamma^{\rm STM}_{\vk,\vk-\vq}=1$) shown in Figure~\ref{fig:figure1}(f) and (g), which reveal that the suppression of backscattering introduced by forbidden time-reversal scattering suppresses the signal at $\vc{q_1}$ [see Figure~\ref{fig:figure1}(g)].
It should be noted that the widely used JDOS and SJDOS approaches use only information about the electronic structure of the host system while neglecting the scattering properties of the impurities except for their ability to conserve or break time-reversal symmetry. In the following sections the analogous results for the improved impurity-specific FT-QPI simulation within the KKR formalism that takes into account the scattering properties of the impurities will be discussed and compared to the standard (S)JDOS approaches. 

\subsection{Impurity specific FT-QPI from first principles \lbl{sec:KKRQPIresults}}

\begin{figure*}
	\centering
	\includegraphics[width=0.80\linewidth]{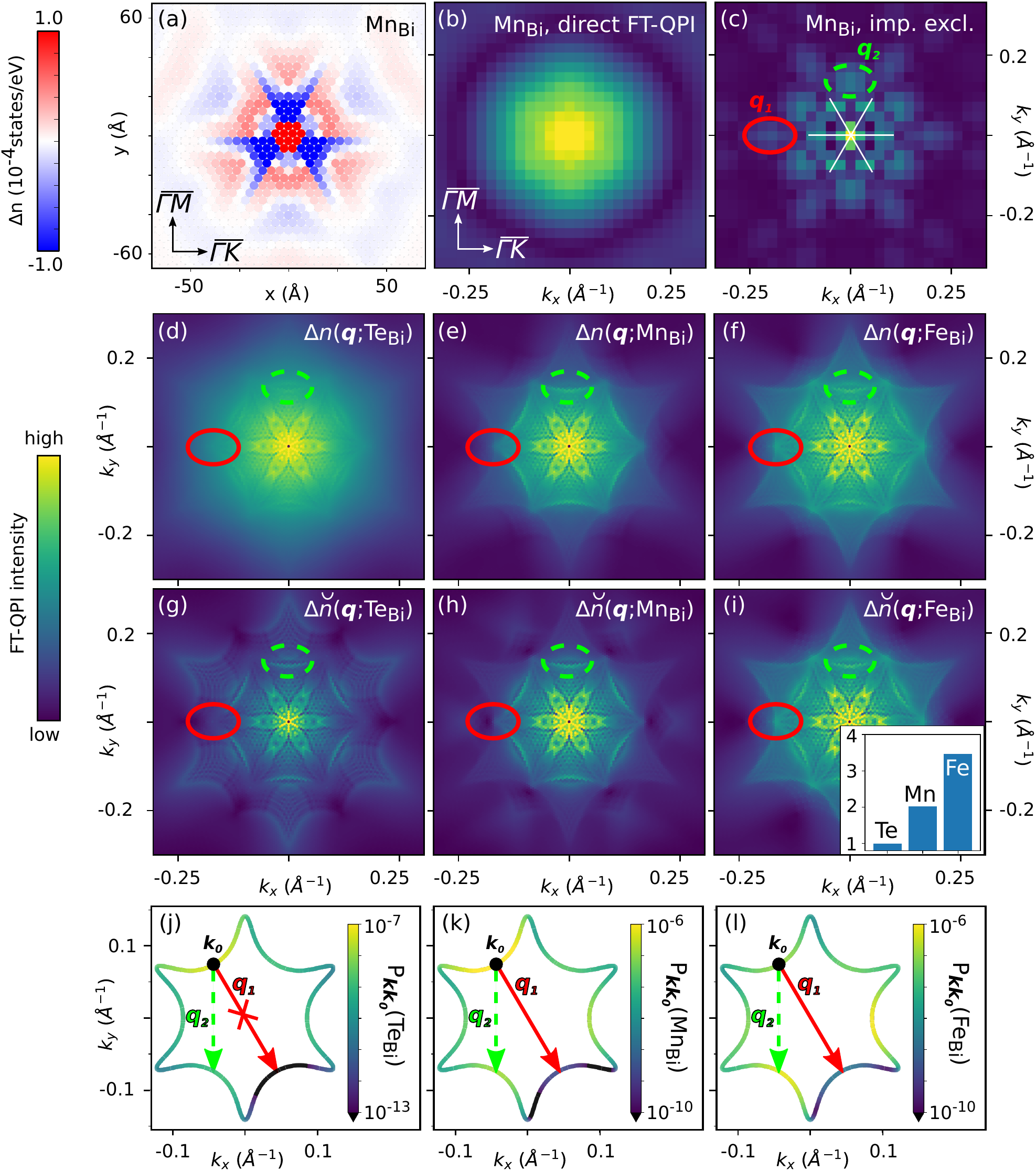}
	\caption{Quasiparticle interference from first principles. (a) Real space image of the charge density oscillations integrated within atomic cells at a distance of $z\approx 1.4 \,\mathrm{\AA}$ above a $\mathrm{Mn}_{\mathrm{Bi}}$ impurity situated at (0,0). (b) FT-QPI of of the real space data shown in (a) computed via FFT. (c) FT-QPI as in (b) but with the near-impurity region within a radius of $R_0\approx10\,\mathrm{\AA}$ (including up to 4th nearest neighbors) set to zero. (d-f) Impurity specific FT-QPI images [after \eq{eq:FTQPI}] for $\mathrm{Te}_{\mathrm{Bi}}$, $\mathrm{Mn}_{\mathrm{Bi}}$ and $\mathrm{Fe}_{\mathrm{Bi}}$ impurities. (g-i) Corresponding FT-QPI images excluding the near-impurity region within $R_0$ [after \eq{eq:denout}] as in (c). 
	(j-l) Scattering rate ($P_{\vc{k},\vc{k_0}}$ in
        $1/(\mathrm{fs}\ \mathrm{at\%}\  \mathrm{d}L)$ with $\mathrm{d}L$ being the Fermi surface line segment) on the Fermi surface for an incoming wave characterized by the wavevector $\vc{k_0}$ for scattering off the $\mathrm{Te}_{\mathrm{Bi}}$, $\mathrm{Mn}_{\mathrm{Bi}}$ and $\mathrm{Fe}_{\mathrm{Bi}}$ impurities, respectively. The inset in (i) shows the ratio of the FT-QPI intensity at $\vc{q_1}$ in (g-i) relative to the signal of the $\mathrm{Te}_{\mathrm{Bi}}$ impurity [$\Delta\breve{n}(\vc{q_1}) / \Delta \breve{n}(\vc{q_1};\mathrm{Te}_\mathrm{Bi})$].
	Image adapted from Ref.~\onlinecite{RuessmannPhD}.}
	\label{fig:figure2}
\end{figure*}

The main calculations of this work are summarized in
Figure~\ref{fig:figure2}. Panel (a) shows the real-space oscillations
in the charge density on the surface, induced by a single subsurface
$\mathrm{Mn}_\mathrm{Bi}$ impurity, often referred to as Friedel
oscillations. The change in the charge density on the surface was
computed for 2148 atomic positions within a radius of
$108\,\mathrm{\AA}$  around the position of the impurity. Note that
supercell-based approaches would have to deal with a size of
approximately $50\times50\times30=75000$ atoms in the unit cell. To
demonstrate the validity of our following simulations we performed a
fast Fourier transform (FFT) of the real-space data, which is shown in
panel (b). The FT-QPI image is dominated by the near-impurity region,
resulting in a strong background centered around $\vc{q}=0$. To reveal
the scattering signatures visible in the long-range tail of the
Friedel oscillations around the impurity, we furthermore substracted
the near-impurity region within a radius of $7.2\,\mathrm{\AA}$ from
the FFT (excluding explicitly the first four shells of neighbors), as
it is frequently done \cite{SessiPV} in QPI analyses of STM
experiments. The result is shown in Figure~\ref{fig:figure2}(c), where
we identify signatures of the two dominant scattering channels,
$\vc{q_1}$ and $\vc{q_2}$, in the resulting image as well as a
star-like feature [highlighted by 6 white lines in
Figure~\ref{fig:figure2}(c)] associated to small angle-scattering off
the impurities. Although this FFT comprises a very large region around
the impurity only a low resolution could be achieved in the resulting
image. It is obvious that for sufficient accuracy much larger regions
need to be included in the FFT, which is, however, numerically expensive.

The results of out new implementation are shown in Figure~\ref{fig:figure2}(d-f) for the full Fourier transform of the $\mathrm{Te}_\mathrm{Bi}$, $\mathrm{Mn}_\mathrm{Bi}$ and $\mathrm{Fe}_\mathrm{Bi}$ impurities, respectively. The images were simulated using a dense $k$-point mesh of $301\times301$ points in the Brillouin zone with a broadening introduced by a small imaginary part in the energy of $\approx 5\,\mathrm{meV}$. The results were checked to be converged with respect to these numerical parameters. The much higher resolution in the full FT-QPI images [see Figure~\ref{fig:figure2}(d-f)] reveal clear evidence of the prominent $\vc{q_1}$- and $\vc{q_2}$-signals, which are far better resolved when the same near-impurity regions including the first four shells around the impurity position is excluded from the Fourier transform [see Figure~\ref{fig:figure2}(g-i)].

Next we compare the different signatures (intensities at $\vc{q_1}$,
$\vc{q_2}$)in the FT-QPI image among the different
impurities. Scattering off the non-magnetic $\mathrm{Te}_\mathrm{Bi}$
defect is characterized by a strong focus in forward (i.e.,
small-angle or $\vc{q}\approx0$) scattering direction. This is a
consequence of the forbidden backscattering that disallows the scattering vector $\vc{q_1}$ as seen by looking at the scattering rate $P_{\vc{k}, \vc{k_0}}=\frac{2\pi}{\hbar}|\Tmat_{\vk,\vc{k_0}}|^2\delta(E_{\vc{k}}-E_{\vc{k_0}})$ \cite{Heers2011,Long2014,Zimmermann2016}, which is illustrated for one particular incoming wave, $\vc{k_0}$, in Figure~\ref{fig:figure2}(j). The strong focus to small angle scattering leads to the absence of a signal at $\vc{q_1}$ and only a moderate intensity at $\vc{q_2}$ compared to the dominant star-like feature around $\vc{q}=0$. In contrast, the intensity at $\vc{q_2}$ is stronger for the $\mathrm{Mn}_\mathrm{Bi}$ and $\mathrm{Fe}_\mathrm{Bi}$ impurities [see Figure~\ref{fig:figure2}(h,i)]. This is a consequence of the lesser focused scattering in forward direction as exemplified in Figure~\ref{fig:figure2}(k,l) for the $\mathrm{Mn}_\mathrm{Bi}$ and $\mathrm{Fe}_\mathrm{Bi}$ impurities. Although the magnetic impurities break the protection against backscattering, the backscattering amplitude is still much smaller than scattering into forward direction or near $120^\circ$-scattering ($\vc{q_2}$). This results in the relatively weak signal of backscattering ($\vc{q_1}$) in the FT-QPI image of single magnetic defects. This result is in line with recent investigations in Mn-doped \BT, where ferromagnetically coupled clusters of atoms were shown to be necessary for the efficient reopening of the backscattering channel within the topological surface state \cite{Sessi16}. 
The higher backscattering rate off the $\mathrm{Fe}_\mathrm{Bi}$ impurity [see the amplitude of $P_{-\vc{k_0},\vc{k_0}}$ in Figure~\ref{fig:figure2}(l)] compared to the $\mathrm{Mn}_\mathrm{Bi}$ defect [see Figure~\ref{fig:figure2}(k)] seems counter intuitive at first sight when considering that Mn has the higher spin moment ($4.48\,\mu_B$) compared to Fe ($3.54\,\mu_B$). However, considering the higher density of states at the Fermi level in the case of the $\mathrm{Fe}_\mathrm{Bi}$ impurity in the context of Wigner's time delay, which relates the higher scattering potential to a longer effective stay at the impurity, the more efficient coupling of the surface state electrons in \BT\ to the $\mathrm{Fe}_\mathrm{Bi}$ impurity becomes apparent. 
As a consequence the FT-QPI intensity at $\vc{q_1}$ in fact increases
by a factor $\approx2$ and $\approx3.5$ [see inset in
Figure~\ref{fig:figure2}(i)], when going from
$\mathrm{Te}_\mathrm{Bi}$  over $\mathrm{Mn}_\mathrm{Bi}$ to the $\mathrm{Fe}_\mathrm{Bi}$  defect, which is a clear signature of the increasing backscattering amplitude.

In summary, our simulations of the impurity specific FT-QPI reveals that not only information on the host's electronic structure can be extracted but also the scattering properties of different impurities are accessible.

\subsection{Comparison to joint density of states approaches \lbl{sec:FTQPIvsJDOS}}

\begin{figure*}
	\centering
	\includegraphics[width=0.75\linewidth]{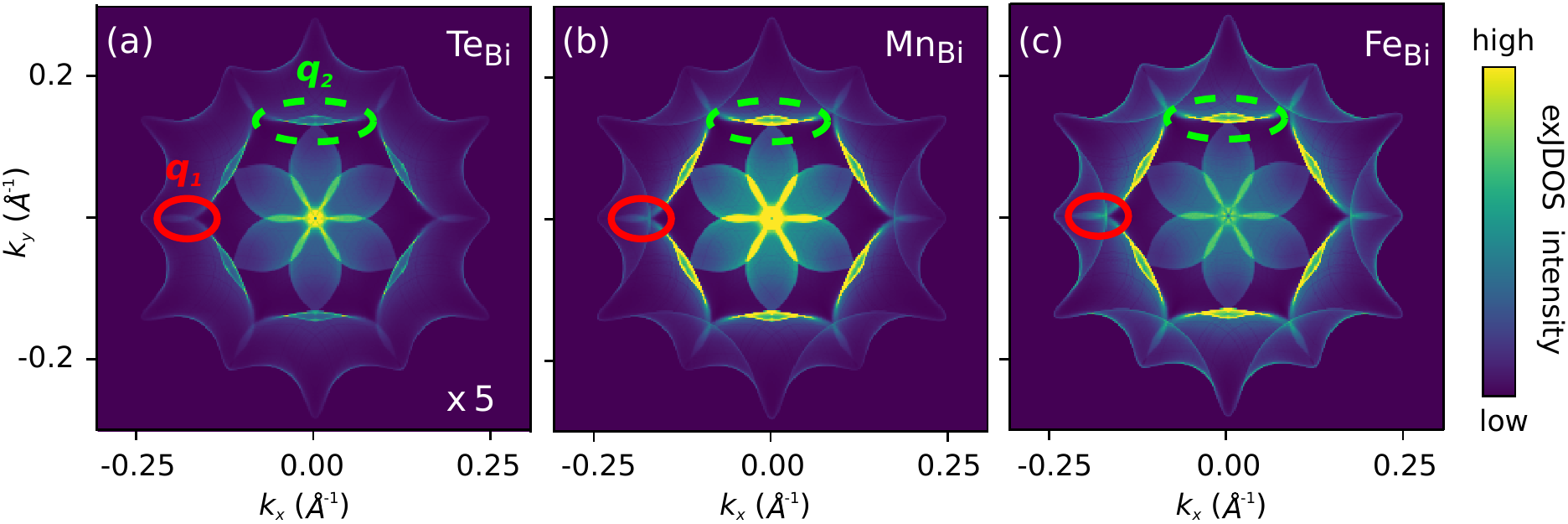}
	\caption{The exJDOS approach for the simulation of FT-QPI images of (a) $\mathrm{Te}_{\mathrm{Bi}}$, (b) $\mathrm{Mn}_{\mathrm{Bi}}$ and (c) $\mathrm{Fe}_{\mathrm{Bi}}$ impurities. A good qualitative agreement with the simulations of $\Delta \breve{n}(\vq;E)$ in Figure~\ref{fig:figure2}(g-i) can be seen, which is a significant improvment over the (S)JDOS images shown in Figure~\ref{fig:figure1}.
	Image adapted from Ref.~\onlinecite{RuessmannPhD}.}
	\label{fig:figure3}
\end{figure*}

Usually the interpretation of experimental FT-QPI images is done by comparison to calculations based on the joint density of states. The comparison between the (S)JDOS in Figure~\ref{fig:figure1}(f,g) on the one hand and the impurity specific results of the FT-QPI presented in Figure~\ref{fig:figure2} on the other hand reveals that a proper description of the impurity scattering is crucial for quantitative understanding of the scattering process off defects. In particular the intensity at the backscattering signal ($\vc{q_1}$) is overestimated in the simple (S)JDOS approaches and the strong focus in small angle scattering is underestimated. A significant improvement was the exJDOS [see \eq{eq:exJDOS}], that does include the correct scattering information of the different impurities and which is shown for the three defects ($\mathrm{Te}_\mathrm{Bi}$, $\mathrm{Mn}_\mathrm{Bi}$ and $\mathrm{Fe}_\mathrm{Bi}$) in Figure~\ref{fig:figure3}. Qualitatively the correct FT-QPI can be reproduced and a strong suppression of the $\vc{q_1}$ signal is found in accordance to the results of Figure~\ref{fig:figure2}. 

This gives an a posteriori justification of the use of the exJDOS
model in the interpretation of experimental QPI images and allows for
the extraction of scattering information as the interplay between the
electronic structure of the host system and the embedded
impurities. It can, however, not be excluded that in other systems,
the terms that are dropped in the JDOS-approaches (e.g.,
mixed $\vc{k}$ contributions or anisotropic scattering rate) become important.

\section{Summary \lbl{sec:Summary}}

We have derived a Green function and \Tmat-matrix based formalism for
the calculation of Fourier-transformed quasiparticle interference
images that are measured with STM. We have implemented our
theory into the
KKR Green function method, allowing for
\textit{ab-initio} calculations for the
translational-invariance-breaking impurity problem. Different than the
simple JDOS models, our approach accounts for the calculated
scattering amplitude of the Bloch wavefunctions off the defects. We have
examined the derivation of the JDOS approaches and shown that they are
not quantitative approximations, but ad hoc qualitative models. Still,
in their simplicity, they comprise an important qualitative part of
the Fourier-transformed QPI physics. We have also derived the
approximation of calculating the Fourier-transformed QPI from a single
defect, compared to the result of multiple, randomly placed defects
and shown that it is reliable at low defect concentrations because of
a cancellation of the multiple-scattering wavefunction phase.

We have applied our KKR-based implementation to non-magnetic and
magnetic defects embedded in the surface of the topological insulator
\BT, providing microscopic insights into the scattering properties
of topological surface state electrons and their response to
time-reversal conserving or breaking defects.

\section*{Acknowledgements}
This work was supported by the Deutsche Forschungsgemeinschaft within
SPP 1666 (Grant No. MA4637/3-1). We furthermore acknowledge financial
support from the VITI project of the Helmholtz Association as well as
computational support from the JARA-HPC Supercomputing Centre at RWTH
Aachen University. 
PR and SB acknowledge support by the Deutsche Forschungsgemeinschaft (DFG, German Research Foundation) under Germany's Excellence Strategy – Cluster of Excellence Matter and Light for Quantum Computing (ML4Q) EXC 2004/1 – 390534769. 
We are indebted to Paolo Sessi and Matthias Bode for illuminating discussions.

\appendix*

\section{Multiple-scattering \Tmat-matrix} 

Here we provide a proof of Equation~(\ref{eq:mstmat}) that avoids 
the infinite-series expansion given in Ref.~\cite{Rodberg}.
The Dyson equation for the \Tmat-matrix can be written in two equivalent forms:
\begin{equation}
\Tmat=\DV + \DV\,\Ghost\,\Tmat = \DV + \Tmat\,\Ghost\,\DV.
\lbl{eq:tmateom}
\end{equation}
We rewrite the first form as
\begin{equation}
\DV=\Tmat\,(1+\Ghost\,\Tmat)^{-1}.
\lbl{eq:dv1}
\end{equation}
Let us denote the impurity sites with the index $n$. It is convenient
to use a notation where we collect the single-site $t$-matrices in a
site-diagonal matrix $\Tmat_{\rm d}$ with $(\Tmat_{\rm
  d})_{nn'}=t_n\,\delta_{nn'}$. In analogy, we collect the diagonal
part of the Green function in the matrix $(\Ghost_{\rm d})_{nn'}=\Ghost_{nn}\,\delta_{nn'}$. $\DV$ is trivially site-diagonal. Then, for the site-diagonal parts we use the second form of Equation~(\ref{eq:tmateom}) that yields $\Tmat_{\rm d} = \DV + \Tmat_{\rm d} \,\Ghost_{\rm d}\DV$, i.e., 
\begin{equation}
\DV = (1+\Tmat_{\rm d}\Ghost_{\rm d})^{-1}\,\Tmat_{\rm d}.
\lbl{eq:dv2}
\end{equation}
Eliminating $\DV$  in (\ref{eq:dv1}) and (\ref{eq:dv2}) we obtain 
\begin{equation}
\Tmat = \Tmat_{\rm d} + \Tmat_{\rm d} (\Ghost - \Ghost_{\rm d}) \Tmat,
\end{equation} 
which is equivalent to (\ref{eq:mstmat}). 


\end{document}